# Electrical Detection of Spin Wave Resonance in a Permalloy Thin Strip

Ziqian Wang, Wei Lu and Bogen Wang

*Abstract*—We investigated the microwave-induced DC response of spin wave resonance (SWR) in a permalloy thin strip via electrical detection. Our experimental results obtained by sweeping the external field reveal that: 1. the amplitude of SWR signals depend on the direction of external field and, 2. unlike the DC response of ferromagnetic resonance, SWR spectra are always anti-symmetrical. The spin dynamics are discussed based on these unusual signals in resonant condition.

## I. Introduction

ELECTRICAL detection of direct current (DC) response of ferromagnetic resonance (FMR) under microwave radiation in very thin sheets of ferromagnetic conductors has been intensively studied since 1960s [1]. The microwave-induced DC response is a non-zero time-averaged signal, which is generated by the microwave induced current and the time-varying anisotropic magnetoresistance (AMR) originates from magnetization precession. This DC response is very informative to further understand the magnetization precessional motion and spin dynamics, e.g., magnetic anisotropy [2], electromagnetic relative phase (the phase difference between electrical field component and magnetic field component of electromagnetic wave) [3] and damping [4]. As a result, DC response detection bears an important role for both scientific points of view and technological perspectives. The motions of spin and magnetization are well described by Landau-Lifshitz-Gilbert (LLG) equation [4, 5]. This equation indicates that, excited by time-varying field of microwave, electron spin will precess about its average direction fixed by a static magnetic field. Under suitable conditions, the amplitude of precession strongly enhanced and is named as "resonance". FMR is a typical resonance for uniform precession mode, while unlike FMR, the spin precession in SWR exhibits as a non-uniform mode in terms of exchange or dipole interaction between spins. SWR usually performs as a mode of standing wave by taking boundary condition into account, and, from a macroscopic view, might result in magnetization precession. SWR is possible to be electrical detected since the time-dependent AMR [6]:

$$\tilde{R} = R_0 + R_A - R_A \sin^2\left[\theta + \Delta\theta\cos(\omega t + \psi)\right]$$

is generated by precessional magnetization. Here $R_0$ represents the resistance while the magnetization $M$ is perpendicular to the microwave induced current, $R_A$ is the decrement resistance, and $\theta$ is the angle of magnetization $M$ with respect to current. $\Delta\theta$ refers to the amplitude of precession angle, $\omega$ is the precession frequency and $\Psi$ is used to identify the phase of magnetization precession. Generally, in the electrical detection, the induced current is originated from microwave and is given by $\tilde{j} = j\cos(\omega t)$. As a result, the voltage response is expressed by Ohm's law as:

$$\begin{aligned}U &= \operatorname{Re}(\tilde{j}) \times \tilde{R} \\ &= j(R_0 + R_A)\cos(\omega t) \\ &\quad - \frac{1}{2}jR_A\Delta\theta\sin(2\theta)\cos(2\omega t + \psi) \\ &\quad - \frac{1}{2}jR_A\Delta\theta\sin(2\theta)\cos\psi\end{aligned} \quad (1)$$

The last time-independent item of (1) is the DC response and can be measured via electrical detection.

In this article, we are focusing on the DC response of exchange-dominated SWR in a permalloy thin strip. The wave vector of this kind of SWR is perpendicular to this strip's plane and is significantly larger than other spin waves, e.g. dipole interaction dominated spin waves. Our purpose is to research the dynamics of this spin wave through electrical detection, and provide a general picture to show how the spin wave evolves in resonant condition.

## II. Experiment: Sample Preparation and Measurement

This work is performed on a $Ni_{80}Fe_{20}$ (permalloy, Py) thin strip with polycrystalline structure. The dimensions of this specimen are: length=2400μm, width=200μm and thickness=49nm (prepared as described elsewhere [7]). The Py thin strip is bonded between two poles. A rectangular waveguide is applied to transmit microwave and ensure them normally propagate into the strip. An electromagnet is employed to provide an external static magnetic field $\mu_0 H_{ex}$ with maximum amplitude of 1.5T. The signal is extracted by using a lock-in amplifier (SR830, Stanford) with the modulation frequency at 8.33 kHz. All data were obtained at

This work was supported in part by the State Key Program for Basic Research of China grants (2007CB613206), the National Natural Science Foundation of China grants (10725418, 10734090, 10990104, and 60976092).

Ziqian Wang is with National Laboratory for Infrared Physics, Shanghai Institute of Technical Physics, Chinese Academy of Sciences, Shanghai, 200083, China. (email: ziqian86@mail.sitp.ac.cn).
Wei Lu is with National Laboratory for Infrared Physics, Shanghai Institute of Technical Physics, Chinese Academy of Sciences, Shanghai, 200083, China. (email: luwei@mail.sitp.ac.cn).
Bogen Wang is with Department of Physics, Nanjing University, Nanjing, Jiangsu Province, 210093, China. (email: bgwang@nju.edu.cn).

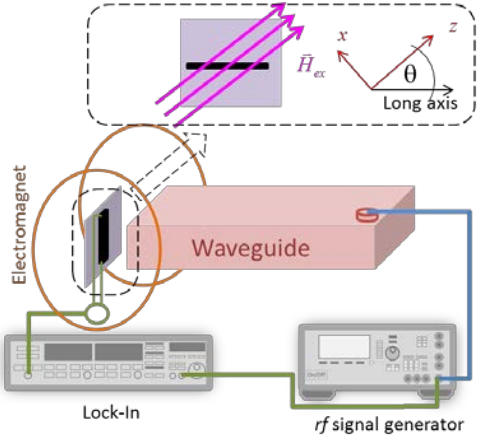

Fig. 1. The schematic diagram of the experimental setup and the selected coordinate system. Here we set the direction of external field $H_{ex}$ as z-axis, and the direction perpendicular to our sample's plane, or xz-plane as y-axis. $H_{ex}$ encloses the angle $\theta$ with respect to the length, or the long-axis of our sample.

room temperature. $H_{ex}$ is in-plane with an angle $\theta$ to the strip's long axis. For simplicity, a coordinate system is selected, as shown in Fig.1.

### III. RESULTS AND DISCUSSION

#### A. Anti-symmetrical SWR DC-response Spectra

The direct comparison between the DC response spectra of SWR and FMR is shown in Fig.2(a). In the previous research [7], researchers found that the FMR DC response represents as a combination of symmetric and anti-symmetric line shapes, shown in Fig.2(b). However, the DC response spectra of SWR are always anti-symmetrical. The ratio of symmetric component and anti-symmetric component reveals the phase difference between precession and microwave-induced AC current. Since the phase difference between precession and magnetic component of microwave can be obtained by solving LLG equation, we can get the information of the relative phase between microwave's electric component and magnetic component. As a result, the electromagnetic relative phase difference cannot be obtained via SWR spectra because it does not consist of symmetric line shape.

The DC-responses measured at different microwave frequencies via sweeping $H_{ex}$ are shown in Fig.2(c). The resonances are associated with FMR and SWR. In spite of the complex physical mechanism of SWR, its anti-symmetrical spectrum is empirically concluded as:

$$U_{DC} \sim \frac{\Delta H (H_{ex} - H_r)}{(H_{ex} - H_r)^2 + \Delta H^2}, \quad (2)$$

here $\Delta H$ represents the line width of SWR signal and $H_r$ is the intensity of external field at resonance. A peak and a nadir exist at each side of $H_{ex} = H_r$.

The anti-symmetrical shape of SWR signal brings facilities for us to find $H_r$ and the signal's amplitude. In this article, we define the amplitude of SWR as the difference between the strengths of signals at peak and nadir, as illustrated in Fig.2(d). According to (1), the amplitude also depends on $\theta$, and is proportional to $\sin(2\theta)$, see in Fig.2(e).

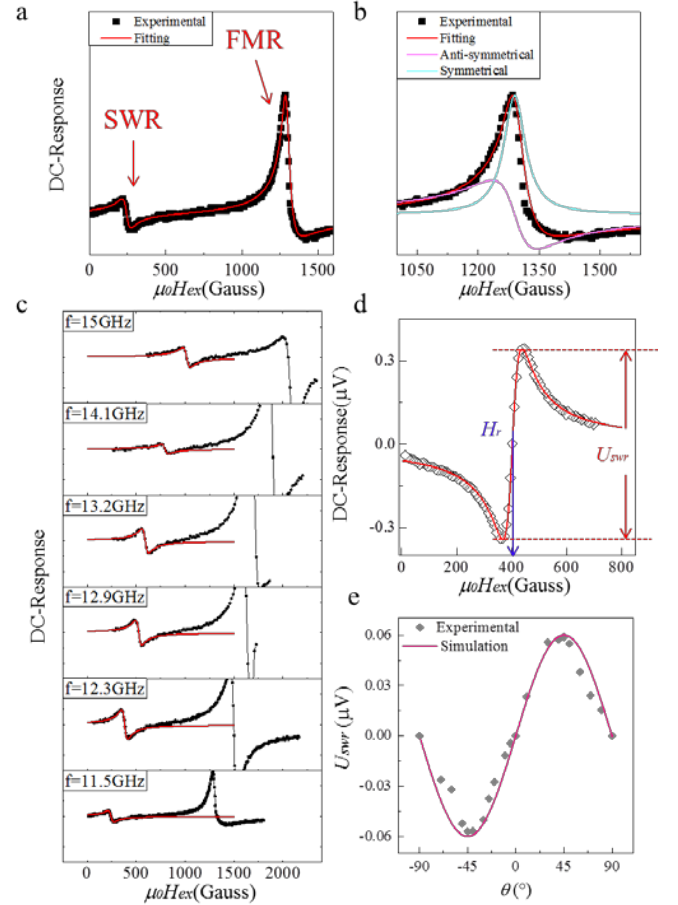

Fig. 2. (a). The comparison between SWR and FMR DC-responses. The frequency of assisted microwave is fixed at 11.5GHz while $\theta$ is 45°. (b). The FMR spectra in (a) exhibits as a typical signal curve, which consists of symmetrical and anti-symmetrical line shapes. (c). The experimental (black dots and lines) and fitted (red lines) of SWR DC-response at different microwave frequencies while $\theta$ is fixed at 45°. (d). The definitions of resonant external field $H_r$ and amplitude $U_{swr}$ for SWR signals in this article. (e). The experimental (grey dots) and fitted (red line) of $\theta$ dependence of $U_{swr}$ at fixed microwave frequency of 10.8GHz.

#### B. The spin dynamics in SWR

The most intrigue issue for SWR is: why the SWR spectra are different from those of FMR? Before this question is answered, we will begin our discussion with the origin of "exchange interaction equivalent magnetic field", or $H_{exchange}$. In our experiment, $H_{exchange}$ is directly obtained as $\mu_0 H_{exchange}$=1100Gauss by recording the dispersion curves for SWR and FMR, see in Fig.3(a). Here we begin the following discussion with calculating the origin of $H_{exchange}$. As noticed in the introduction, the exchange interaction dominated spin wave in thin strip exhibits as a standing spin wave mode (SSW) due to the pinning condition at the sample's surfaces, accordingly we may draw a picture for SSW, shown in Fig.3(b) and Fig.3(c).

For the $n^{st}$ spin in a SSW chain, considering the exchange interaction, the LLG equation is modified as [8]:

$$\dot{\vec{M}}_n = \gamma \tilde{H} \times \vec{M}_n - \frac{\alpha}{|\vec{M}|}\dot{\vec{M}}_n \times \vec{M}_n + \gamma\lambda\left(\vec{M}_{n-1} + \vec{M}_{n+1}\right)\times \vec{M}_n \quad (3)$$

here $M_n$ represents the n$^{st}$ spin's magnetic momentum, $\gamma$ is the electron's gyromagnetic ratio, $\alpha$ represents the damping coefficient and $\lambda$ is the exchange coefficient. Only the exchange interactions of nearby spins are considered. For simplicity, we treat the amplitude of each spin's precessional motion follows sinusoidal distribution, and the precession for the spin at antinode of SSW is

$$\vec{M}_{sw} = \left(e^{i\omega t}m_{sw\_x}, \quad e^{i\left(\omega t - \frac{\pi}{2}\right)}m_{sw\_y}, \quad M\right)$$

for small $m_{sw\_x}$ and $m_{sw\_y}$, and $M_n$ is written by:

$$\vec{M}_n = \left(e^{i\omega t}m_{x(n)}, \quad e^{i\left(\omega t - \frac{\pi}{2}\right)}m_{y(n)}, \quad M\right)$$

$$= \left(e^{i\omega t}m_{sw\_x}\sin\frac{n\pi}{N}, \quad e^{i\left(\omega t - \frac{\pi}{2}\right)}m_{sw\_y}\sin\frac{n\pi}{N}, \quad M\right)$$

And we obtain:

$$\gamma\lambda\left(\vec{M}_{n-1} + \vec{M}_{n+1}\right)\times \vec{M}_n = 2\gamma\lambda\sin\frac{n\pi}{N}\left(\cos\frac{\pi}{N}-1\right)$$

$$\times \left(e^{i\left(\omega t - \frac{\pi}{2}\right)}m_{sw\_y}M, \quad e^{i\omega t}m_{sw\_x}M, \quad 0\right)$$

So the exchange interaction of nearby spins acts on the n$^{st}$ spin is equivalent to a static field $H_{exchange}$ and is given by:

$$\vec{H}_{exchange} = \left(0, \quad 0, \quad 2\lambda M\left(1-\cos\frac{\pi}{N}\right)\right) \quad (4)$$

(4) implies that, for series of spins with same phases, precession traces and different amplitudes, the exchange interaction among them can be equalized as a static magnetic field with the same direction as $H_{ex}$. If the precessional motions for each spin in SSW follow other distributions, $H_{exchange}$ for each spin might different. However, $H_{exchange}$ is also equivalent to a static field without time-varying component if each spin precesses at the same phase.

Nonetheless, it is indicated by solving LLG equation that, for two spins, while the damping coefficients, the static magnetic fields (including $H_{exchange}$ and $H_{ex}$) and the exciting magnetic fields $h_{mw}$ are the same, the precessional amplitude $\Delta\theta$ and phase $\Psi$ for them are also the same. Furthermore, if the static magnetic fields are different while $\alpha$ and the assisted time varying magnetic field $h_{mw}$ are the same, their precessional phases will be different. In non-resonant condition, the phase of spin precession with respect to $h_{mw}$ is approximately 0 (for $H_{ex}$ larger than $H_r$) or 180° (for $H_{ex}$ smaller than $H_r$), and $\Delta\theta$ is very small. That is to say, in non-resonant condition, the presumption, each spin in a SSW precesses coherently, is correct. While $H_{ex}$ is swept at resonant region, with the increased amplitude of spin precession, the non-uniformity of spin precession inside the ferromagnetic sheet is strengthened, accompany with the emergence of phase difference between two nearby spins along y-axis. Consequently, from a macroscopic point of view, the time-varying component of magnetoresistance is offset by the phase differences between spins, and equals zero. That is why the SWR signals are 0 at the expected resonant field $H_{swr}$.

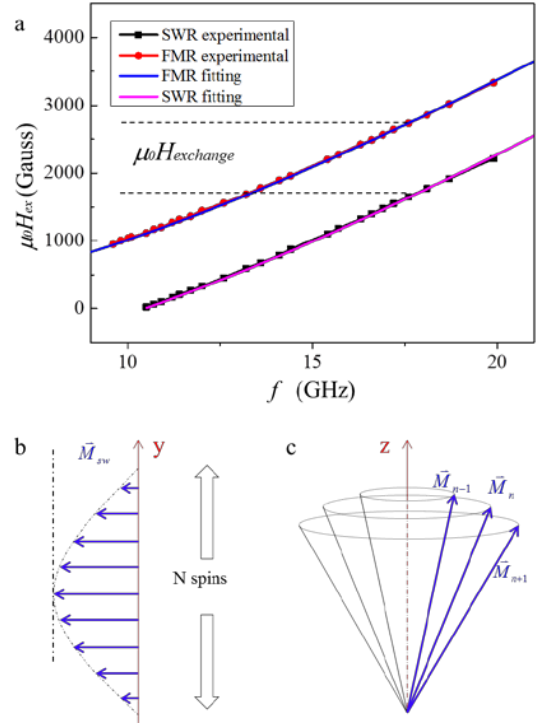

Fig. 3. (a). The experimental (colored dots) and fitted (colored lines) of SWR and FMR dispersion curves. $H_{exchange}$ is indicated. (b). The amplitude of the precessional motions for each spin follows sinusoidal distribution, while the magnetic momentum of spin at the antinode of SSW (consists of $N$ spins) is recorded as $M_{sw}$. (c). Only nearby spins, $M_{n-1}$ and $M_{n+1}$ are taken into account while calculating $H_{exchange}$ for a certain $M_n$.

If the DC response of SWR for $H_{ex}$ smaller than $H_r$ is written by

$$U_{DC}\left(H_{ex} < H_r\right) = -\frac{1}{2}jR_A\Delta\theta\sin(2\theta)\cos\psi,$$

in the field-swept mode, the signal for $H_{ex}$ larger than $H_r$ should be expressed as:

$$U_{DC}\left(H_{ex} > H_r\right) = -\frac{1}{2}jR_A\Delta\theta\sin(2\theta)\cos(\psi - \pi)$$

$$= \frac{1}{2}jR_A\Delta\theta\sin(2\theta)\cos\psi \quad (5)$$

$$= -U_{DC}\left(H_{ex} < H_r\right)$$

That is to say, the DC response of SWR is anti-symmetrical with respect to $H_{ex}=H_r$.

IV. CONCLUSIONS

We have demonstrated the anti-symmetrical SWR DC response spectra in a Py thin strip through electrical detection. The amplitude of SWR DC response is found to be influenced by the direction of $H_{ex}$. We have also obtained $H_{exchange}$ via measuring the dispersion curves of FMR and SWR. The origin of $H_{exchange}$ is discussed. Furthermore, we have found that in the resonant region, it is impossible for spins inside our sample precess as a standing wave mode with the same phase due to the requirement of LLG equation, and the DC response thereby vanishes at $H_{ex}=H_r$. In the non-resonant condition, the precessional phase of spin wave changes 180° from $H_{ex}<H_r$ to $H_{ex}>H_r$, and finally results in the anti-symmetrical line shape of SWR DC response.